\documentstyle[honnef,pslatex,psfig]{article}  
\frompage{000} \topage{000}                                              
\def\rightmark{Hydrodynamic Models for heavy-ion collisions, and beyond} 
\def\leftmark{A.\ Dumitru}
 
\title{Hydrodynamic Models for Heavy-Ion Collisions, and beyond} 
\authors{
{A.\ Dumitru$^{1,a}$, J.\ Brachmann$^{2}$, E.S.\ Fraga$^3$,
W.\ Greiner$^{2}$, A.D.\ Jackson$^4$, J.T.\ Lenaghan$^4$,
O.\ Scavenius$^4$, H.\ St\"ocker$^{2}$ %
}\\[2.812mm]
{\normalsize
\hspace*{-8pt}$^1$ Physics Dept., Columbia Univ., 
538 W.\ 120th Street, New York, NY 10027, USA\\
\hspace*{-8pt}$^2$ Institut f\"ur Theor.\ Physik, J.W.\ Goethe
Univ., Robert-Mayer Str.\ 10, 60054 Frankfurt, Germany\\
\hspace*{-8pt}$^3$ 
Brookhaven National Laboratory,
Upton, NY 11973, USA\\
\hspace*{-8pt}$^4$ The Niels Bohr Institute,
Blegdamsvej 17, DK-2100 Copenhagen {\O}, Denmark
}}
 
\abstract{A generic property of a first-order phase transition
{\em in equilibrium}, and in the limit of large entropy per unit of conserved
charge, is the smallness of the isentropic speed of sound in the
``mixed phase''. A specific prediction is that this should lead to a
non-isotropic momentum distribution of nucleons in the reaction plane (for
energies $\sim40A$~GeV in our model calculation). On the other hand, we show
that from present effective theories for low-energy QCD one does not expect the
thermal transition rate between various states of the effective potential to be
much larger than the expansion rate, questioning the applicability of the
idealized Maxwell/Gibbs construction. Experimental data could soon provide
essential information on the dynamics of the phase transition.}
 
\begin{document}
 
\maketitle
\vspace*{24pt}

\section{Introduction}
Heavy-ion collisions at high energy are devoted mainly to the study of
strongly interacting matter at high temperature and density. In particular,
a major focus is the search for observables related to restoration of
chiral symmetry and deconfinement, as predicted by lattice QCD to occur at
high temperature, and in thermodynamical equilibrium~\cite{lattice}.
Also, various effective models for low-energy QCD predict
chiral symmetry restoration at high net baryon density and low
temperature~\cite{largemu}.
 
Several observables for the transition to deconfined matter have been
proposed, such as electromagnetic radiation, strangeness enhancement and
equilibration, charmonium dissociation, and ``irregularities'' in the
hydrodynamic flow pattern~\cite{qm99}. The latter, in particular, represents
an observable that is related to the effective potential at finite
temperature and density, because hydrodynamical expansion is driven by
pressure gradients, and the pressure is minus the value of the
effective potential at the global minimum. Therefore, if indeed a new minimum
of the effective potential opens up at some temperature or density, the
pressure gradients should change and allow observation of that new state.
Here, we discuss the differential distribution of the momenta of nucleons in
the reaction plane (spanned by the beam axis and the impact parameter axis)
as an observable for the emergence of a second
minimum of the effective potential, which defines a new thermodynamical
state of hot matter. Finally, we shall discuss if a phase transition close
to equilibrium is likely to be a reasonable approximation, how one can
improve on that model, and possible consequences for observables.

\section{Effective Potential for the Chiral Phase Transition}
For illustration, consider the following Lagrangian for the approximately
$O(4)$
symmetric chiral field $\Phi=(\sigma,{\bf \pi})$ coupled to (constituent)
quarks $q=(u,d)$:
\begin{equation}
{\cal L} =
 \overline{q}[i\gamma ^{\mu}\partial _{\mu}-g(\sigma +i\gamma _{5}
 {\bf \tau} \cdot {\bf \pi} )]q
+ \frac{1}{2}(\partial _{\mu}\sigma \partial ^{\mu}\sigma + \partial _{\mu}
{\bf \pi} \partial ^{\mu}{\bf \pi} )
- U(\sigma ,{\bf \pi})\quad.
\label{sigma}
\end{equation}
The potential, exhibiting both spontaneously and explicitly broken 
chiral symmetry, is
\begin{equation} \label{T=0_potential}
U(\sigma ,{\bf \pi} )=\frac{\lambda ^{2}}{4}(\sigma ^{2}+{\bf \pi} ^{2} -
{\it v}^{2})^{2}-H_q\sigma\quad.
\end{equation}
The vacuum expectation values of the condensates are $\langle\sigma\rangle 
={\it f}_{\pi}$ and $\langle{\bf \pi}\rangle =0$, where ${\it f}_{\pi}=93$~MeV 
is the pion decay constant. The explicit symmetry breaking term is due to the 
finite current-quark masses and is determined by the PCAC relation which gives 
$H_q=f_{\pi}m_{\pi}^{2}$, where $m_{\pi}=138$~MeV is the pion mass.  
This leads to $v^{2}=f^{2}_{\pi}-{m^{2}_{\pi}}/{\lambda ^{2}}$.  The value of 
$\lambda^2 = 20$ (used throughout this paper) leads to a $\sigma$-mass, 
$m^2_\sigma=2 \lambda^{2}f^{2}_{\pi}+m^{2}_{\pi}$, equal to 600~MeV.
For $g>0$, the finite-temperature one-loop effective potential also 
includes the following contribution from the quarks:
\begin{equation}\label{T>0_potential}
V_q(\Phi) = d_q T
\int \frac{{\rm d}^3k}{(2\pi)^3} \log\left(1+e^{-E/T}\right)\quad.
\end{equation}
Here, $d_q=24$ denotes the color-spin-isospin-baryon charge degeneracy of the 
quarks. $V_q(\Phi)$ depends on the order parameter field $\Phi$ through 
the effective mass of the quarks entering
the expression for the energy $E \equiv \sqrt{k^2 + g^2 \Phi^2}$.
The quarks constitute the 
heat bath in which the long-wavelength modes of the chiral field, i.e.\ the 
order parameter, evolve, resulting in the finite-temperature
effective potential $V_{\rm eff} \equiv U+V_q$.
\begin{figure}[htb]
                 \insertplot{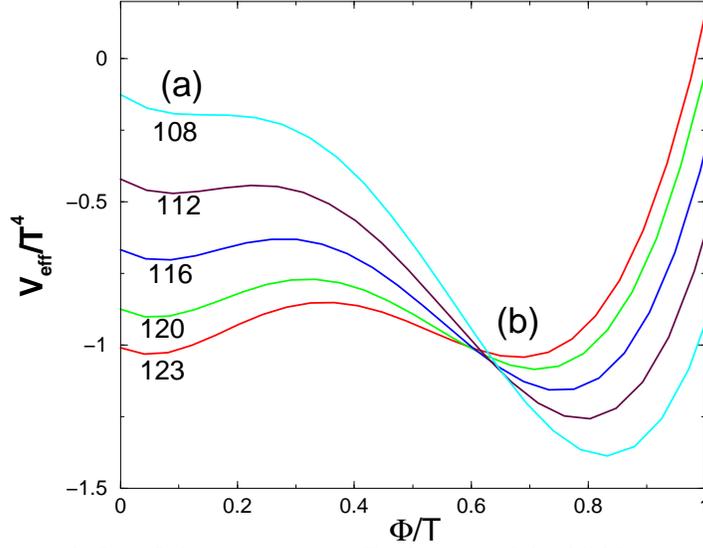}
\vspace*{-1.1cm}
\caption[]{An example for a finite-temperature effective potential
exhibiting two states (a) and (b), respectively, as a function of the
order parameter field $\Phi$ in $\sigma$-direction
and temperature $T$, for vanishing net baryon charge. See text and
ref.~\protect\cite{Scavenius:2000bb} for details.
}
\vspace*{-.6cm}
\label{pot1d}
\end{figure}
For rather large values of $g$,
the theory exhibits a first-order (chiral symmetry restoring) phase
transition. For example, for $g=5.5$ the critical temperature is
$T_c=123.7$~MeV at vanishing baryon-chemical potential $\mu$.
At the phase transition line in the $T-\mu$ plane, 
$V_{\rm eff}$ exhibits two degenerate minima labeled (a) and (b) in
Fig.~\ref{pot1d}, respectively. The first is the thermodynamical state
corresponding to restored chiral symmetry, while the second corresponds to
broken symmetry and smoothly approaches the physical vacuum at $T=0$.
The two states are separated by a barrier which becomes smaller as the
temperature decreases, ending in a point of inflection at $T_{sp}
\approx108$~MeV (for our set of parameters and $\mu=0$), the so-called spinodal
instability. At this temperature, there is no more barrier to the true
ground state and the order parameter ``rolls'' down towards the state (b),
a process called spinodal decomposition. In thermodynamical equilibrium,
the order parameter is localized in either one of the two minima, whichever
is at lower energy. At the phase boundary, where the two states are degenerate,
the system is in the ``mixed phase'', the expectation value of $\Phi$ being
peaked both about $\Phi_{(a)}$ as well as $\Phi_{(b)}$. A Maxwell construction
can be performed two obtain the relative probabilities for the system to be
in either of the two states.

If one is interested in the hydrodynamical evolution of a system in local
equilibrium, and described by the effective potential $V_{\rm eff}$, all one
needs to know is the value of $V_{\rm eff}$ at its global minimum, which is
the value of the grand canonical potential (density) at the given temperature
and chemical potential, i.e.\ minus the pressure.
Once the function $p(T,\mu)$ is known, the evolution (in local equilibrium)
of some initial condition is uniquely determined by the equations of ideal
relativistic hydrodynamics.

\section{Hydrodynamics of Strongly Interacting Matter}  
Hydrodynamics is defined by (local) energy-momentum and net charge
conservation~\cite{Landau},
\begin{equation} \label{Hydro}
\partial_\mu T^{\mu\nu}=0 \quad,\quad
\partial_\mu N_i^{\mu}=0\quad.
\end{equation}
$T^{\mu\nu}$ denotes the energy-momentum tensor, and $N_i^{\mu}$ the net
four-current of the $i$th conserved charge. We will explicitly consider only
one such conserved charge, namely the net baryon number. We implicitly assume
that all other charges which are conserved on strong-interaction time scales
vanish locally. The corresponding four-currents are therefore identically zero,
cf.\ eq.~(\ref{idfluid}), and the conservation equations are trivial.

For ideal fluids, the energy-momentum tensor and the net baryon current
assume the simple form~\cite{Landau}
\begin{equation} \label{idfluid}
T^{\mu\nu}=\left(\epsilon+p\right) u^\mu u^\nu -p g^{\mu\nu}
\quad,\quad
N_B^{\mu}=\rho u^\mu \quad,
\end{equation}
where $\epsilon$, $p$, $\rho$ are energy density, pressure, and net baryon
density in the local rest frame of the fluid, which is defined by
$N_B^\mu=(\rho,{\bf 0})$. $g^{\mu\nu}={\rm diag}(+,-,-,-)$ is the metric
tensor, and $u^\mu=\gamma(1,{\bf v})$ the four-velocity of the fluid
(${\bf v}$ is the three-velocity and $\gamma=(1-{\bf v}^2)^{-1/2}$ the
Lorentz factor). The system of partial differential
equations~(\ref{Hydro}) is closed by choosing an equation of state (EoS)
in the form $p=p(\epsilon,\rho)$.

The EoS employed in this section exhibits a first order phase transition to a
Quark-Gluon Plasma (QGP). The hadronic phase consists of nucleons
interacting via relativistic scalar and vector fields~\cite{GorEOS};
the pressure is obtained from the one-loop effective potential for the
nucleons, similar to eq.~(\ref{T>0_potential}),
plus that of free thermal pions. The QGP phase is described within
the framework of the MIT-Bag model as an ideal gas of $u$ and $d$ quarks
and gluons, with a bag para\-meter $B^{1/4}=235$~MeV, resulting in a critical
temperature $T_c\simeq 170$~MeV at $\rho=0$, while the critical baryon-chemical
potential is $\mu_c=1.8$~GeV at $T=0$. The first
order phase transition is constructed via
Gibbs' conditions of phase equilibrium. Thus, by construction the two states
of the effective potential are coexisting in equilibrium: as the fluid
expands, an increasing fraction of matter is transfered from the
high-temperature state (a) to the low-temperature state (b). Supposedly,
that happens through nucleation of bubbles of hadronic matter within the
QGP matter~\cite{kapusta,Scavenius:2000bb}.
If that process occurs arbitrarily close to
equilibrium, the two thermodynamical states remain degenerate throughout the
phase transition region, such that the pressure is the same in both states.

\subsection{The effect of a first-order equilibrium phase transition
on directed flow}
The equilibrium (first-order) transition discussed above has interesting
implications regarding the hydrodynamical expansion pattern of the hot and
dense matter~\cite{Rischke:1996nq}.
Expansion of a perfect fluid conserves the entropy current
$s^\mu=su^\mu$, as follows from eqs.~(\ref{Hydro}) upon
contraction with the contravariant four-velocity $u_\nu$, and from the
thermodynamical identities ${\rm d}\epsilon=T{\rm d}s+\mu {\rm d}\rho$,
$\epsilon=Ts-p+\mu\rho$. Thus, the entropy
density current $s^\mu$ is proportional to the baryon density
current $N_B^{\mu}$; each fluid element traces a path of $s/\rho={\rm const.}$
in the phase diagram. The pressure gradient among neighbouring fluid
elements, which is responsible for the acceleration in a given spatial
direction, is given by $\nabla p = c_s^2 \nabla\epsilon$, where
\begin{equation}
c_s^2 \equiv \frac{\partial p}{\partial\epsilon}\Bigr|_{s/\rho={\rm const.}}
\end{equation}
denotes the isentropic speed of sound. 
\begin{figure}[htb]
\vspace*{-1.1cm}
                 \insertplot{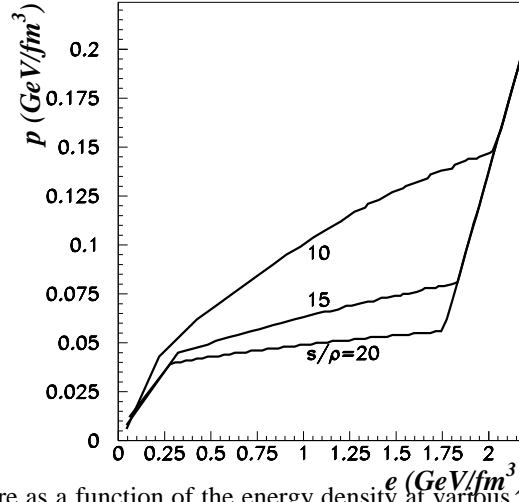}
\vspace*{-1.1cm}
\caption[]{The pressure as a function of the energy density at various
values for the specific entropy (for the EoS described in the text).
}
\vspace*{-.5cm}
\label{eps_n}
\end{figure}
Fig.~\ref{eps_n} shows the pressure as a function of the energy density at
various values for the specific entropy $s/\rho$. (The ``wiggles'' in the
curves are due to the plotting procedure and do not reflect the numerical
accuracy, which is much better.) One clearly observes a rapid increase
of $p(\epsilon)$ within the hadronic phase, followed by a less rapid
increase in the coexistence (``mixed'') phase, and finally the transition to
pure quark-gluon matter at very high energy density. In particular, the
hotter the fluid, that is the larger $s/\rho$, the more does $p(\epsilon)$
flatten out in the mixed phase. For $s/\rho\rightarrow\infty$ we obtain
$p={\rm const.}$ inbetween the two pure phases. That behavior can be
understood immediately as following from Gibbs' conditions of phase
equilibrium.
$s/\rho\rightarrow\infty$ implies $\mu/T\rightarrow0$, and thus in this limit
the temperature is the only remaining intensive thermodynamical variable the
effective potential (and the pressure) can depend on.
Now, in phase equilibrium $T=T_c={\rm const.}$ and so the pressure
$p(T_c)\equiv p_c$ is constant as well.
In other words, the isentropic speed of sound
$c_s^2$ must be small during the time when a fluid element's trajectory
through the phase diagram coincides with the phase coexistence line (in the
$T-\mu$ plane). 

If that prediction based on the equilibrium phase diagram is indeed relevant
for collisions of heavy ions, it should be visible in the excitation function
of the so-called directed in-plane flow~\cite{Rischke:1996nq,1f_dirflow},
$\langle
p_x/N\rangle$. That observable is defined as the average momentum per nucleon
in impact parameter ($x-$) direction~\cite{Danielewicz:1985hn}, and due to
the kinematics of the collision is
particularly sensitive to the early stage of the reaction, where the phase
transition might occur. A more refined observable is the distribution
of momentum in the reaction plane, which is proportional to the
triple-differential cross section for the process $AB\rightarrow p+X$,
${\rm d}\sigma/{\rm d}^3p$. To zeroth approximation, it can be obtained by
integrating the baryon current $N_B^{\mu}$ over the freeze-out hypersurface
of the nucleons, $\Sigma^\mu$:
\begin{equation}
\frac{{\rm d}\sigma}{{\rm d}^3p} 
\propto \int {\rm d}\Sigma_\mu N_B^\mu \delta\left(
m_N{\bf u}-{\bf p}\right)\quad.
\end{equation}
In what follows we assumed for simplicity that freeze-out, or at least
onset of strong dissipative correction to the perfect-fluid energy-momentum
tensor, occurs on an equal-time hypersurface in the center of mass frame,
${\rm d}\Sigma^\mu=({\rm d}^3x,{\bf 0})$. Furthermore, we integrated the
differential
cross section over the momentum component $p_y$, perpendicular to both the
impact parameter vector and the beam axis.

\begin{figure}[htb]
\vspace*{-1.1cm}
\insertplot{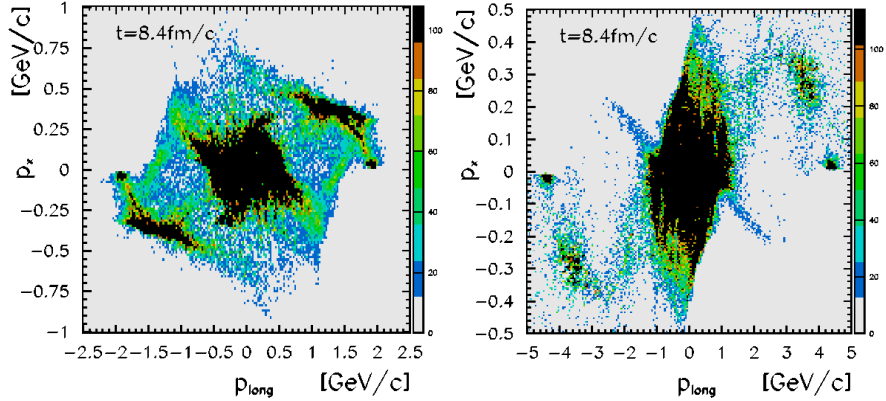}
\vspace*{-2.1cm}
\caption[]{Distribution of the time-like component of the net-baryon
four-current in momentum space ($p_x-p_{z}$ plane).
Pb+Pb collisions at ${b=3}$~fm, 
$E^{kin}_{Lab}=8A$~GeV (left frame) and $40A$~GeV (right frame), respectively.
}
\label{p_x}
\end{figure}
Typical distributions of nucleons in momentum space are depicted in
Fig.~\ref{p_x}~\cite{jb_flowmax}. The left frame corresponds to
relatively low collision energy and specific entropy $s/\rho\approx10$.
Therefore, $c_s^2$ is not small in that case even if the energy density at the
center exceeds $\sim0.5$~GeV/fm$^3$, and the state (a) of the effective
potential should be populated according to the equilibrium EoS (see
Fig~\ref{eps_n}).
So, since $c_s^2$ is not small, (energy-) density gradients in coordinate space
reflect in non-vanishing pressure gradients $\nabla p$, which in turn
act to make the distribution of nucleons in momentum space more or less
isotropic. The {\em net} momentum in $x$-direction at any fixed longitudinal
momentum $p_z$ is obviously rather small, as positive and negative
contributions largely cancel~\cite{Brachmann:2000xt}.

On the other hand, Fig.~\ref{p_x} also shows that the structure of the flow
changes qualitatively at higher energy, $E^{kin}_{Lab}\approx40A$~GeV, where
on average $s/\rho\approx20$. The fact that $c_s^2$ is now smaller than
at lower energy, by a factor of two in the present model, actually
{\em prevents} more isotropic redistribution of the baryon number in
momentum-space~\cite{jb_flowmax}. The distribution is clearly very different
from that at the lower energy, with almost no
nucleons in the upper left or bottom right quadrants where
$p_x\cdot p_{z}<0$~\cite{jb_flowmax}.

\section{The nucleation rate, and supercooling down to the spinodal}
The Gibbs/Maxwell construction relies on the
assumption that the thermal barrier penetration rate is much larger than
the local expansion rate. In a slowly expanding 
system, the phase transition would proceed through the nucleation of bubbles 
of the ``true vacuum'' state via thermal
activation~\cite{kapusta,Scavenius:2000bb}.
The nucleation rate per unit volume per unit time is 
expressed as 
\begin{equation} \label{decrate}
\Gamma={\cal P}~ e^{-F_b/T}\quad,
\end{equation}
where $F_b$ is the free energy of a critical bubble and where the
prefactor ${\cal P}$ provides a measure of the saddle point of the
Euclidean action in functional space. Our
analysis will concentrate on the exponential barrier penetration 
factor, and we will approximate ${\cal P}$ by $T^4$, which is obtained
from dimensional considerations at $\mu=0$.
In order to determine the role of bubble nucleation in the evolving system 
and to be able to compute the decay rate $\Gamma$, it is necessary to 
study the critical bubble and some of its features.  The critical 
bubble is a radially symmetric, static solution of the Euler-Lagrange field 
equations that satisfies the boundary condition $\Phi (r \to \infty) 
\longrightarrow \Phi_{(a)}$.
Energetically, this boundary condition corresponds to an 
exact balance between volume and surface contributions which defines the 
critical radius $R=R_c$.  The critical bubble is unstable with respect to 
small changes of its radius.  For $R < R_c$, the surface energy 
dominates, and the bubble shrinks into the false vacuum.  For $R>R_c$, the 
volume energy dominates, and the bubble grows driving the decay process.

The critical bubble can be found by minimizing the free energy
\begin{equation}
F_b(\Phi,T)=4\pi\int \, r^2\, {\rm d}r\, \left[\frac{1}{2}\left(\frac{{\rm d}
\Phi}{{\rm d}r}\right)^2
+V_{\rm eff}(\Phi,T)\right]\quad,
\label{free}
\end{equation}
with respect to the field $\Phi$. This can be performed numerically without
further approximation~\cite{Scavenius:2000bb}.
\begin{figure}[htb]
\vspace*{-.9cm}
\insertplot{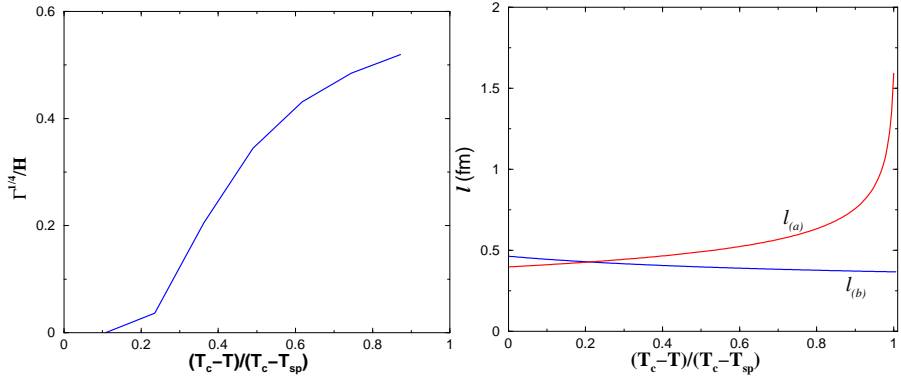}
\vspace*{-2.4cm}
\caption[]{$\Gamma^{1/4}$ divided by an assumed expansion rate of $H = 
1$~fm$^{-1}$ as a function of temperature (left frame).
Correlation length of the chiral order parameter in
state (a) and in state (b). The latter corresponds to an equilibrium
transition.}
\vspace*{-.5cm}
\label{figdecrate}
\end{figure}
Fig.~\ref{figdecrate} shows the resulting decay rate, divided by an assumed
expansion rate of $H=1$~fm$^{-1}$, which represents a rough estimate for the
expansion rate of the three-volume of a comoving fluid element, ${\rm d}V\equiv
{\rm d}\Sigma\cdot u$, in high-energy collisions. $\Gamma^{1/4}$ specifies the
inverse spatial and temporal scale of thermal field fluctuations into the
broken symmetry state. Clearly, for the above effective field theory
$\Gamma^{1/4}$ is not much larger than $H$, even if one takes into account that
both factors are only known approximately. 
While the results of Fig.~\ref{figdecrate} do not provide 
an unambiguous prediction for the fate of the supercooled state, they do make 
it appear likely that the rapidly expanding system has an appreciable 
probability of remaining in the restored symmetry phase even close to the 
spinodal instability. Thus, at least some fraction of all heavy ion events 
should show traces of this non-equilibrium transition.
The familiar idealized Gibbs/Maxwell construction of 
equilibrium thermodynamics may not be appropriate for the description of 
phase transitions in high-energy heavy ion collisions. If so, the anisotropy
of the differential nucleon cross section in the reaction plane, predicted
in the previous section on the basis of the equilibrium phase diagram,
should be {\em absent}. Rather, the small net momentum in impact parameter
direction (around $p_z\sim0$), as observed experimentally~\cite{E895}
at the top BNL-AGS energy $E^{kin}_{Lab}\approx10A$~GeV, should remain
essentially zero or diminish even further. That is because the phase
equilibrium with small
velocity of sound $c_s^2$ does not occur in a rapid out of equilibrium
transition, where the order parameter remains localized in state (a) of
the effective potential down to the spinodal.

Instead, other experimental observables of the phase transition may emerge.
Consider the behavior of the correlation length of the $\Phi$-field.
It is obvious from Fig.~\ref{pot1d} that the curvatures of the effective 
potential at the minima $\Phi_{(a)}$ and $\Phi_{(b)}$ are rather different for 
$T<T_c$. Thus, the effective mass, $m_{\rm eff}^2={\rm d}^2 V_{\rm eff}/ 
{\rm d}\Phi^2$, and the correlation length, $l=1/m_{\rm eff}$, of the field 
will also be different. In particular, if the rapidly expanding system
supercools appreciably and if the order parameter is trapped in the
metastable state, one can expect a clear increase of $l$.
The right frame of Fig.~\ref{figdecrate} shows $l$ in the states (a) and (b),
respectively,
as a function of the degree of supercooling.  The correlation length at 
(a) increases as that minimum disappears.  By contrast, 
there is a smooth decrease in the correlation length at the true global
minimum, corresponding to an equilibrium transition.
In the present model $l_{(a)}$ can exceed $l_{(b)}$ by
as much as a factor of $2-3$, depending on the degree of supercooling.
The precise values of $l$ depend on the specific effective model 
adopted, but the qualitative observation that $l_{(a)}$ increases as the
system approaches the spinodal instability is general.

\section{Conclusions}
We discussed generic properties of the effective potential for a first-order
phase transition in equilibrium, in particular the vanishing of the
isentropic velocity of sound in the limit of very large entropy per
unit of conserved charge (baryon charge in our case). As an example for
how such properties of the effective potential show up in the dynamics of
hot and dense QCD matter we discussed the so-called directed flow of
nucleons, i.e.\ the differential cross section in the reaction plane.
Furthermore, we showed that the forthcoming results of the Pb+Pb
reactions at $E^{kin}_{Lab}=40A$~GeV can test the applicability of the picture
of hot QCD-matter as a heat-bath with small isentropic speed of sound
(mixed phase) to heavy-ion collisions
in that energy domain. {\em If} it holds true, our model calculations
predict an increase of the directed net in-plane momentum
as compared to top BNL-AGS energy, and a
nonisotropic momentum distribution around midrapidity.
During the assumed first-order equilibrium phase transition
pressure gradients along isentropes are too small to 
work towards a more isotropic momentum distribution.

The experimental test of the equilibrium phase transition picture would
also provide valuable constraints on effective field theories for low-energy
QCD. As mentioned above, for the present model thermal barrier penetration
from state (a) to state (b), and vice versa, is not materially larger than
rough estimates for the expected expansion rates, thus leading to the
possibility of significant supercooling, close to the spinodal instability.

The experimental observation of supercooling effects and spinodal 
decomposition would also be important as a matter of principle.  Ideally, 
signatures of phase transitions should be order parameter related and 
should reveal properties of the equilibrium phase diagram.
While the spinodal instability is not part of the equilibrium 
phase diagram, it is rather close.  Familiar experiments in condensed 
matter physics on a wealth of hysteresis phenomena (i.e.\ the analogue of 
superheating and supercooling) make it clear that spinodal instabilities 
can be studied on ``macroscopic'' time scales~\cite{spin}.
The spinodal instability, 
if found, would certainly be among the most direct information which 
relativistic heavy ion collisions can provide regarding the QCD phase 
transition.

\section*{Acknowledgements}
A.D.\ thanks the organizers for the invitation to attend and present this
work, and also acknowledges support from DOE grant, Contract No.\
DE-FG-02-93ER-40764. E.S.F.\ is partially supported by the U.S.\
Department of Energy under Contract No.\ DE-AC02-98CH10886 and by 
CNPq (Brazil) through a post-doctoral fellowship. J.B.\, W.G.\, and H.S.\
acknowledge partial support by BMBF, DFG, and GSI.
 
\section*{Notes}
\begin{notes}
\item[a]
E-mail: dumitru@nt3.phys.columbia.edu
\end{notes}

\vfill\eject
\end{document}